\documentclass[aps,prb,reprint]{revtex4-1}
\usepackage{lipsum}
\usepackage{amssymb,amsmath}
\usepackage{subfig}
\usepackage{graphicx}
\usepackage{color}
\usepackage{hyperref}
\usepackage{listings}
\usepackage{fancybox}
\usepackage{caption}
\makeatletter
\newcommand*{\balancecolsandclearpage}{%
  \close@column@grid
  \cleardoublepage
  \twocolumngrid
}

\makeatother
\begin{document}

\title{	Fodis: software for protein unfolding analysis}
\author{Nicola Galvanetto$^{1,a, \ast}$}
\author{Andrea Perissinotto$^{1, a}$}
\author{Andrea Pedroni$^{1, 4}$}
\author{Vincent Torre$^{1,2,3,b}$}

\affiliation{$^{1}$International School for Advanced Studies (SISSA), via Bonomea 265, Trieste 34136, Italy 
	$^{2}$Cixi Institute of Biomedical Engineering, Ningbo Institute of Materials Technology and Engineering, Chinese Academy of Sciences, Zhejiang 315201, P. R. China 
	$^{3}$Center of Systems Medicine, Chinese Academy of Medical Sciences, Suzhou Institute of Systems Medicine, Suzhou Industrial Park, Suzhou, Jiangsu, 215123 P.R. China 
	$^{4}$Present Address: Department of Neuroscience, Karolinska Institutet, 171 77, Stockholm, Sweden 
	$^{a}$Equal contribution 
	$^{b}$Senior Author \\
	$^{\ast}$Corresponding author, Email: \texttt{nicola.galvanetto@sissa.it}}

\date{Accepted for publication in Biophysical Journal February 07, 2018; DOI: \url{https://doi.org/10.1016/j.bpj.2018.02.004}}

\begin{abstract}
	The folding dynamics of proteins at the single molecule level has been studied with single-molecule force spectroscopy (SMFS) experiments for twenty years, but a common standardized method for the analysis of the collected data and for sharing among the scientific community members is still not available.
	We have developed a new open source tool --Fodis-- for the analysis of Force-distance curves obtained in SMFS experiments, providing an almost automatic processing, analysis and classification of the obtained data. Our method provides also a classification of the possible unfolding pathways and structural heterogeneity, present during the unfolding of proteins.\\
	\textbf{Availability}: The source code, executables for Win, Mac OS and Linux, datasets and user guide are available for download at 
	\url{https://github.com/nicolagalvanetto/Fodis}.
	
\end{abstract}

\maketitle

\section{Introduction}
The investigation of proteins has been greatly advanced from the recent technical improvements of cryo-EM, which have allowed the determination of the structure of many proteins avoiding the bottleneck of crystallization (1). However, fundamental single protein properties, such as their folding/unfolding dynamics and structural heterogeneity have to be addressed with other experimental methods like single molecule optical methods (sm-fluorescence and smFRET) or Single-Molecule Force Spectroscopy (SMFS) (2). SMFS is an application of AFM and of optical tweezers (OT; Fig.1A, B) in which the force F required to unbind a ligand or unfold a polymer is measured as a function of the length d, at pN and nm resolution. The obtained Force-distance (F-d) curves are composed by a series of force peaks characterizing the unfolding of structural segments. A force peak occurs during the unfolding of an $\alpha$-helix and/or $\beta$-sheets followed by an unfolded segment. When a protein changes its conformation, the force peaks of F-d curves change their location and amplitude: from these changes, structural information of the molecular rearrangements is obtained. The location of a force peak is obtained by fitting the experimental F-d curve with the Worm Like Chain (WLC) model, providing the values of the contour lengths (Lc). From the value of Lc, it is possible to estimate the number of a.a. of the polypeptide unfolded between the occurrence of consecutive force peaks, and to probe its structural heterogeneity from the distribution of the force peaks at different pulling rates (3).
Recent improvements allow the collection of more data and of an higher quality (4), but there is no universal ready-to-use and data sharing platform for the analysis of obtained F-d curves. Several algorithms to reduce manual intervention have been published (5–7), but the analysis has not been fully automated yet and still requires a significant amount of decisions to be taken by the researcher.
We developed Fodis (for “Force-distance software” Fig. 1C), an open-source software providing an all-in-one environment with several tools for the analysis of F-d curves, from the raw curves to data representations and novel analytical methods.

\begin{figure}[h]
	\centering
	\includegraphics[width=0.5\linewidth]{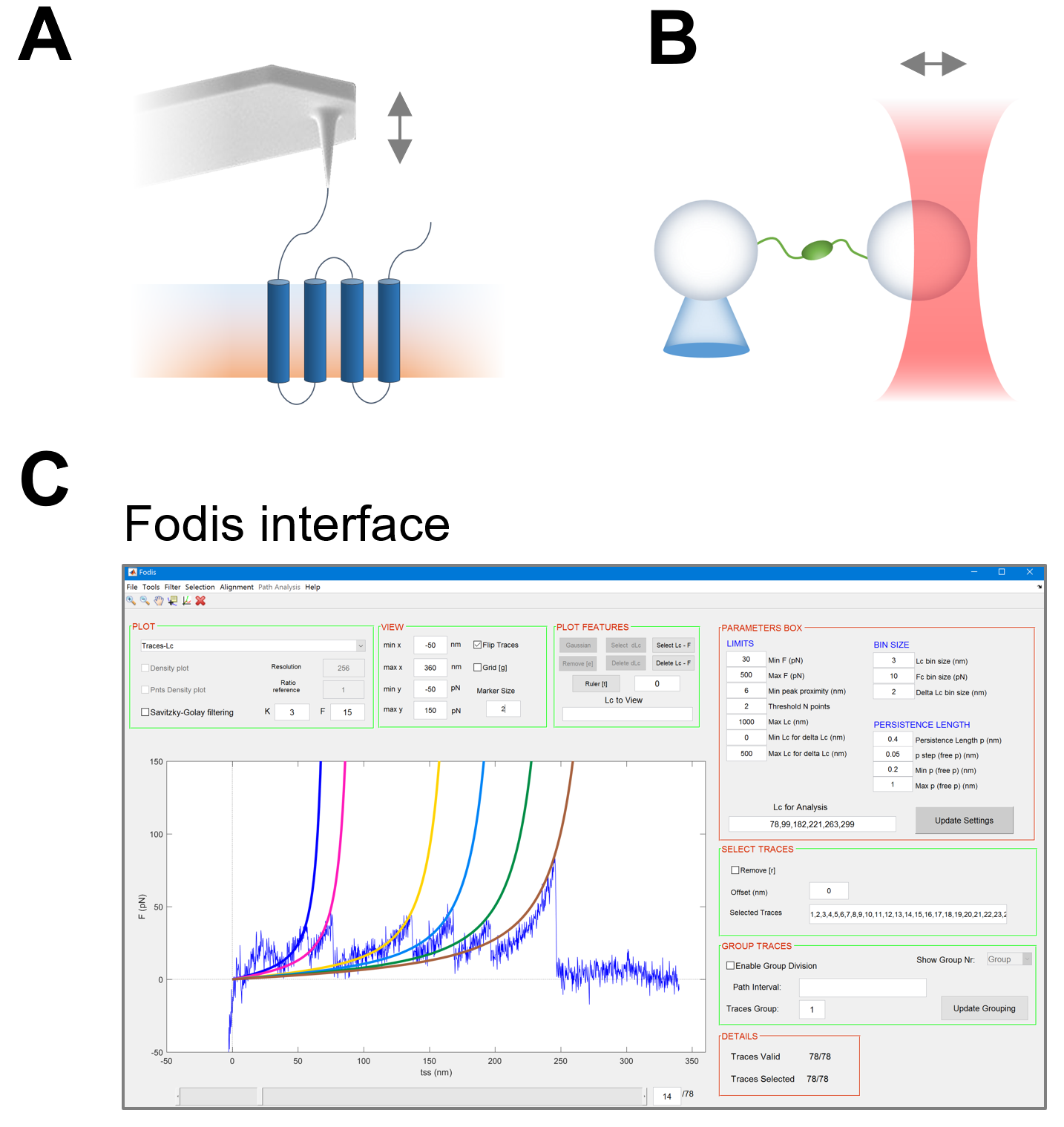}
	\captionsetup{justification=raggedright,singlelinecheck=false}
	\caption{Cartoon of an SMFS experiment based on (A) AFM and (B) optical tweezer. (C) example of the graphic user interface of Fodis. }
	\label{fig:cpu}
\end{figure}

\begin{figure*}
	
	\centering
	\includegraphics[width=0.6\linewidth]{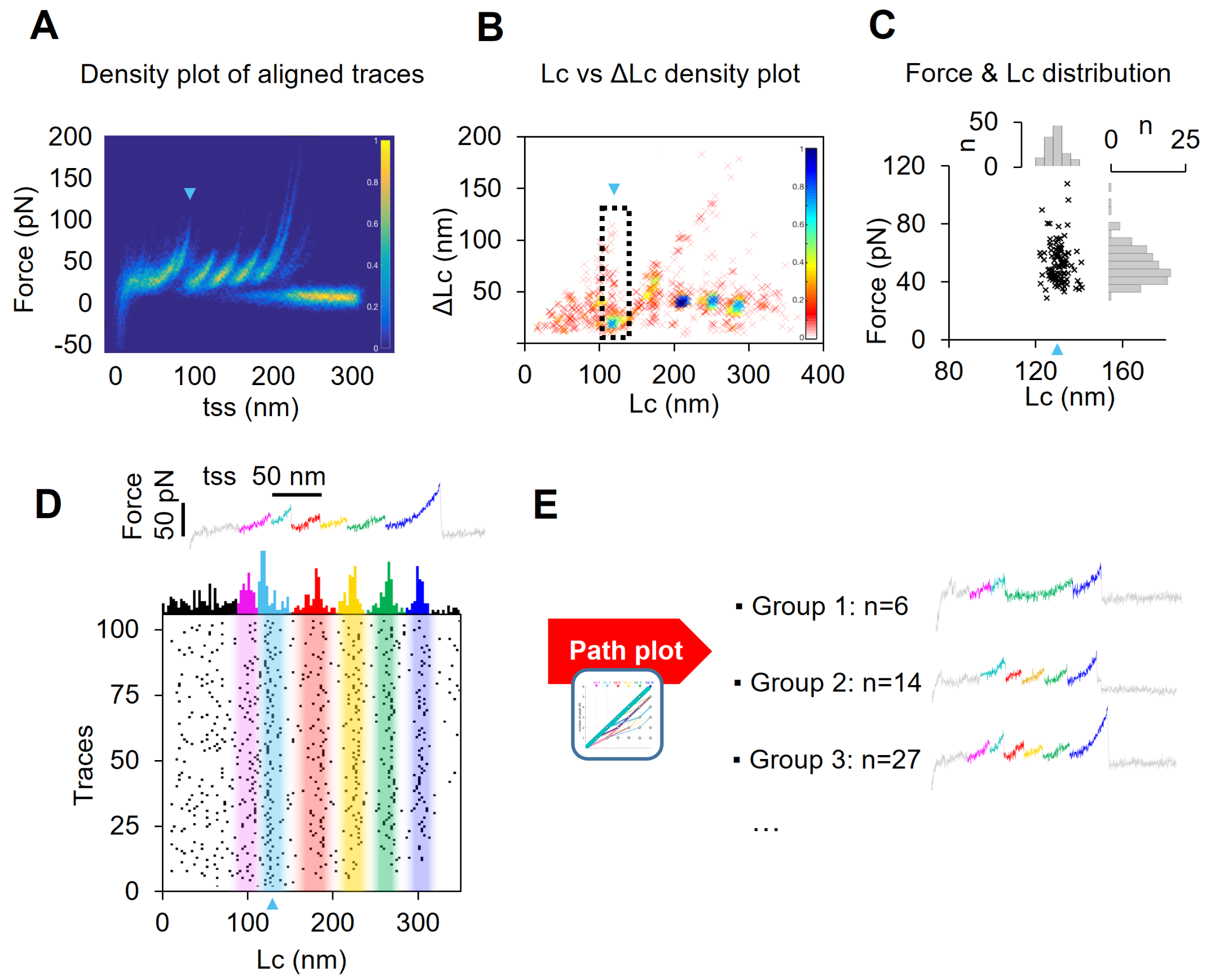}
	\captionsetup{justification=raggedright,singlelinecheck=false}
	\caption{(A) Superimposition of 104 F-d curves as a density plot, after the automatic alignment (Supplementary Fig. 3). (B) Scatter plot of all detected peaks from the population of the 104 curves. The x-dimension is the contour length (Lc) and the y-dimension reports the difference between the considered peak and the previous force peak (ΔLc). Point density is color coded according to the bar on the right. (C) Scatter plot of the unfolding events between 120 and 140 nm (triangle in A) and the histograms of force and Lc distribution. (D) Top, a representative F-d curve. Bottom, cumulative histogram that counts the number of peaks along Lc coordinate and the Global Matrix (a N×B matrix where N is the number of traces (rows) while B is the number of bins along the Lc coordinate. Detected peaks are here plotted as black pixels). Colors identify the different unfolding groups. (E) The Path plot algorithm (Figure S5 and S6) splits the population of curves in homogeneous groups representing different unfolding pathways.}
	\label{fig:networks}
\end{figure*}

\section{Materials and Methods}

Fodis filters the F-d curves according to few set parameters (Fig. S1 and S2 in the Supporting Material), that consist mainly in the expected distance of the fully stretched molecule (i.e. contour length maximum or Lc max). Fodis aligns automatically filtered F-d curves (Fig. 2A and Fig. S3) and provides a thorough statistical characterization of their variability (Fig. 2B, C; Fig. S4).  Filtered F-d curves –with a similar value of Lc max–  are heterogeneous, but Fodis identifies clusters of similar F-d curves with similar force peaks. These clusters, depending on the experiment under consideration, correspond either to the unfolding of different molecules, or to the same molecule in different states, or to different unfolding pathways (Fig. 2E; Fig. S5). Fodis can open raw data files from different AFM manufacturers (JPK and Bruker Instruments), and also ASCII matrices of F-d values (Supplementary Note 1). The software is designed to assist the user in the most common/critical steps with a reversible workflow allowing to keep track of the original information. Fodis permits to save working sessions, to export curves selections and to extract graphical data representations.

\section{Results and Discussion}
As a testbed, we processed F-d curves obtained from SMFS experiments on oocyte membranes after the overexpression of the cyclic nucleotide-gated CNGA1 channels (8) (Supplementary data). On a typical notebook computer (4 GB memory, 2.0 GHz CPU), Fodis loads 1000 F-d curves in 5 seconds and perform a filtering routine in less than 20 seconds. Filtering depends on the type of analyzed data and in Fig S2 results from one-day experiment are presented. F-d curves selected from several experiments can be automatically aligned for statistical analysis: for this purpose we implemented the method developed by Bosshart et al. (7) with some modifications (see Fig. S3 and Supplementary note 3). All F-d curves are coded as strings into a Global Matrix (GM), a binary representation of the position of detected force peaks (see Figure 2D). From the GM, different statistical tests and graphical representations are generated and users can integrate new and ad hoc analyses and tests. The Global Histogram of Maxima (GHM, Fig. S4) counts the frequency of occurrence of force peaks with a value of Lc in a given range. The GHM can be fitted by a multi Gaussian distribution to identify the most probable Lc positions of unfolding events. Fodis creates cumulative scatter plots of peak populations that can be selected on the basis of their position (Lc) or force, and then individually analyzed to extract force and position distributions (Fig. 1B, C).  The Lc- ΔLc plot shown in Fig. 1B is an example of how we can extract information about the periodicity of peak occurrence from the Global Matrix (Figure 2D).
Fodis allows also the identification of different clusters of F-d curves (Path Plot: Fig. 1E; Fig. S5 and S6), by coding F-d curves on the basis of the number of force peaks and of their corresponding value of Lc. The Path Plot algorithm is used for a graphical representation of different unfolding pathways, generalizing previous methods for unfolding pathways determination and selection presented by Schönfelder et al. (9), Yu et al. (2) or Thoma et al (10).

\section{Conclusions}

In Fodis we implemented relevant published and new algorithms for SMFS analysis. Fodis provides a toolbox to the SMFS community for the development and sharing of analytical methods in an all-in-one open-source software. It has been developed to meet the needs of a broad audience, both for researchers with no programming skills and for those who could contribute to Fodis’ future versions.

\subsection*{Software Availability}
Fodis is available under the  Apache License, version 2.0. Source code, executables, datasets and full documentation are available for download in GitHub at \url{https://github.com/nicolagalvanetto/Fodis}. The latest release, version 1.2, is archived in Zenodo: DOI: 10.5281/zenodo.841277.

\subsection*{Author Contributions}
N.G. conceived the project, developed the algorithms, helped in writing the code and wrote the paper. A. Perissinotto developed the algorithms and wrote the code. A. Pedroni helped in developing the software. V.T. conceived the project and helped in writing the paper. All the authors read and approved the final article.

\subsection*{Acknowledgments}

The authors tank Dr Sourav Maity for data collection and Paolo Fabris for the initial technical support.  The authors thank Manuela Lough for checking the English. The authors thank Dr Michele Allegra and Dr Giampaolo Zuccheri for useful discussions. V.T. shall recognize the support from 3315 project of Ningbo/Cixi, Zhejiang province, China.

\bibliography{bib}

\subsection*{References}
1. 	Bai, X., G. McMullan, and S.H.W. Scheres. 2015. How cryo-EM is revolutionizing structural biology. Trends in Biochemical Sciences. 40: 49–57.

2. 	Yu, H., M.G.W. Siewny, D.T. Edwards, A.W. Sanders, and T.T. Perkins. 2017. Hidden dynamics in the unfolding of individual bacteriorhodopsin proteins. Science. 355: 945–950.

3. 	Hinczewski, M., C. Hyeon, and D. Thirumalai. 2016. Directly measuring single-molecule heterogeneity using force spectroscopy. PNAS. 113: E3852–E3861.

4. 	Otten, M., W. Ott, M.A. Jobst, L.F. Milles, T. Verdorfer, D.A. Pippig, M.A. Nash, and H.E. Gaub. 2014. From genes to protein mechanics on a chip. Nat Meth. 11: 1127–1130.

5. 	Marsico, A., D. Labudde, T. Sapra, D.J. Muller, and M. Schroeder. 2007. A novel pattern recognition algorithm to classify membrane protein unfolding pathways with high-throughput single-molecule force spectroscopy. Bioinformatics. 23: e231–e236.

6. 	Puchner, E.M., G. Franzen, M. Gautel, and H.E. Gaub. 2008. Comparing Proteins by Their Unfolding Pattern. Biophysical Journal. 95: 426–434.

7. 	Bosshart, P.D., P.L.T.M. Frederix, and A. Engel. 2012. Reference-Free Alignment and Sorting of Single-Molecule Force Spectroscopy Data. Biophysical Journal. 102: 2202–2211.

8. 	Maity, S., M. Mazzolini, M. Arcangeletti, A. Valbuena, P. Fabris, M. Lazzarino, and V. Torre. 2015. Conformational rearrangements in the transmembrane domain of CNGA1 channels revealed by single-molecule force spectroscopy. Nature Communications. 6: 7093.

9. 	Schönfelder, J., R. Perez-Jimenez, and V. Muñoz. 2016. A simple two-state protein unfolds mechanically via multiple heterogeneous pathways at single-molecule resolution. Nature Communications. 7: ncomms11777.

10. 	Thoma, J., N. Ritzmann, D. Wolf, E. Mulvihill, S. Hiller, and D.J. Müller. 2017. Maltoporin LamB Unfolds β Hairpins along Mechanical Stress-Dependent Unfolding Pathways. Structure. 25: 1139–1144.e2.

\balancecolsandclearpage

\begin{figure*}
	
	\centering
	\includegraphics[width=0.6\linewidth]{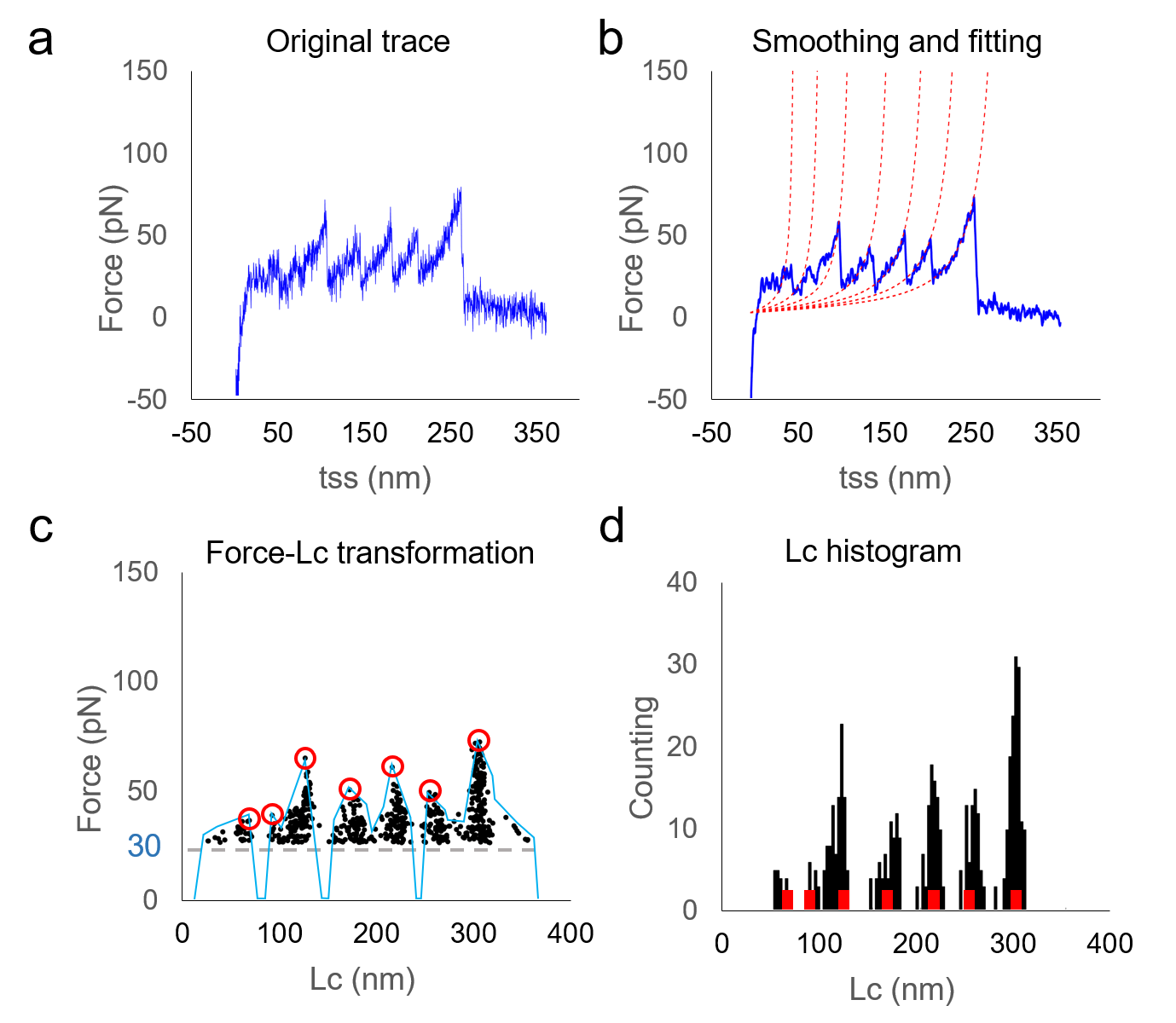}
	\captionsetup{justification=raggedright,singlelinecheck=false}.
	\makeatletter
	\renewcommand{\fnum@figure}{\figurename~S1}
	\makeatother
	\caption{\textbf{Preprocessing of F-d curves}:  (a) Example of a Force-distance curve (F-d) obtained by a commercial AFM (JPK), representing the relation between the force exerted by the cantilever tip vs tip-sample-separation (tss). (b) the same curve after  smoothing with Savitzky-Golay (1) filter, which preserves discontinuity. This filter reduces noise enhancing peaks detection. Dotted red lines: fit of the rising phases of the curve with the Worm-like chain (WLC) model (2) (fixed persistence length p=0.4 nm and free contour length (Lc), see equation (1) in Supplementary Note 2 for more details). (c) Transformation of the F-d curve in the (F,Lc) plane using equation (2) in Supplementary Note 2. Only points with a value of F larger than 30 pN were considered. The cyan line is the Force Profile, and it is used for the automatic detection of the peaks (see Supplementary note 2). (d) Histogram of the values of Lc  shown in (c), also called “barrier position histogram”(3). Red bars represent the position of detected peaks. }
	
\end{figure*}

\begin{figure*}
	
	\centering
	\includegraphics[width=0.9\linewidth]{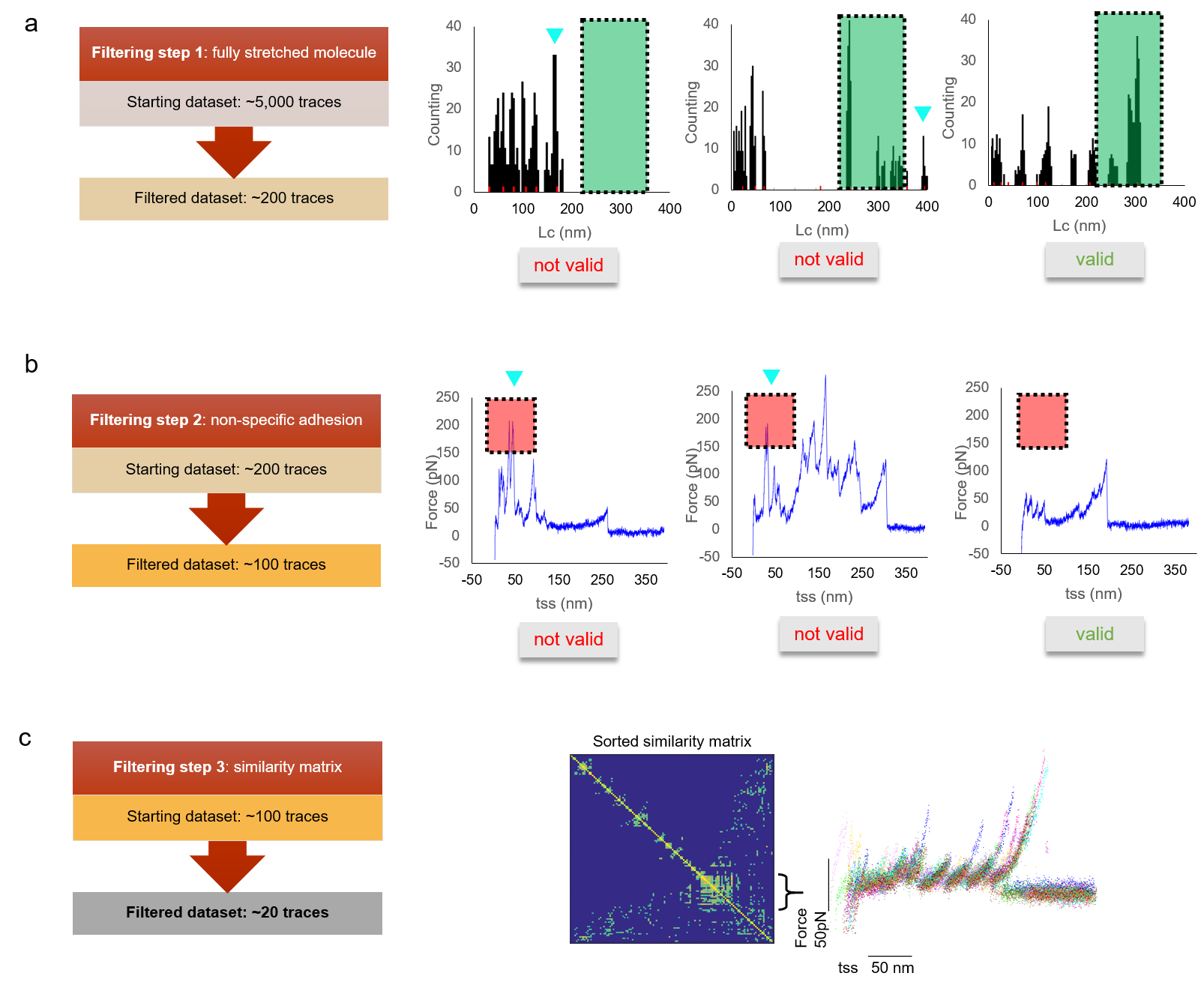}
	\captionsetup{justification=raggedright,singlelinecheck=false}.
	\makeatletter
	\renewcommand{\fnum@figure}{\figurename~S2}
	\makeatother
	\caption{\textbf{Automatic filtering of F-d curves}:  (a) starting from a dataset of ~10,000 curves (Supplementary data), 3 steps are used to find the appropriate curves.  Filtering step 1 tests the fully stretched condition (i.e. the total length of the primary structure of the protein). This condition excludes curves that are longer or shorter than expected value of Lc. The length of the trace corresponds to the position of the last peak in the contour length histogram (Fig S1 d). The expected length of a protein is equal to the number of amino acids (N of a.a.) times the single a.a. length (~0.4 nm (4)). The measured length, estimated in the in the contour length histogram, highly depends on the value of persistence length (p; commonly set to 0.4 nm (5)), but which can vary between 0.3 nm and 0.8 nm (4). Because of all these uncertainties, we suggest to use a filtering window centered in the value of the expected length, and large at least ±30\% of that value. The expected length for the CNG channel is ~280 nm (690 a.a. x 0.4 nm). The green window in (a) is 210-360 nm and determines the Lc region within which the trace must end. The filtering step 1 is intended to be a coarse tool for excluding curves with a wrong length.
		(b) filtering step 2 discards curves that display high non-specific adhesion at the very beginning of the unfolding  (curves with forces over 150 pN in the first 70 nm were discarded (6)). 
		(c) filtering step 3 finds groups of similar curves. We define similarity as the value of cross correlation between the contour length histograms (Fig. S1 d) of two curves (see Supplementary note 3, equation 3). If N is the number of curves, a symmetric N x N similarity matrix is formed containing all the correlation values for each couple of curves. The resulting matrix is ordered with the symmetric approximate minimum permutation algorithm(7) (symamd MATLAB function) to obtain clusters of similar curves. Those clusters can be opened, inspected, and compared with a candidate trace or with control experiments. More details about the similarity matrix can be found in the user guide of the software.
		}
	
\end{figure*}

\begin{figure*}
	
	\centering
	\includegraphics[width=0.8\linewidth]{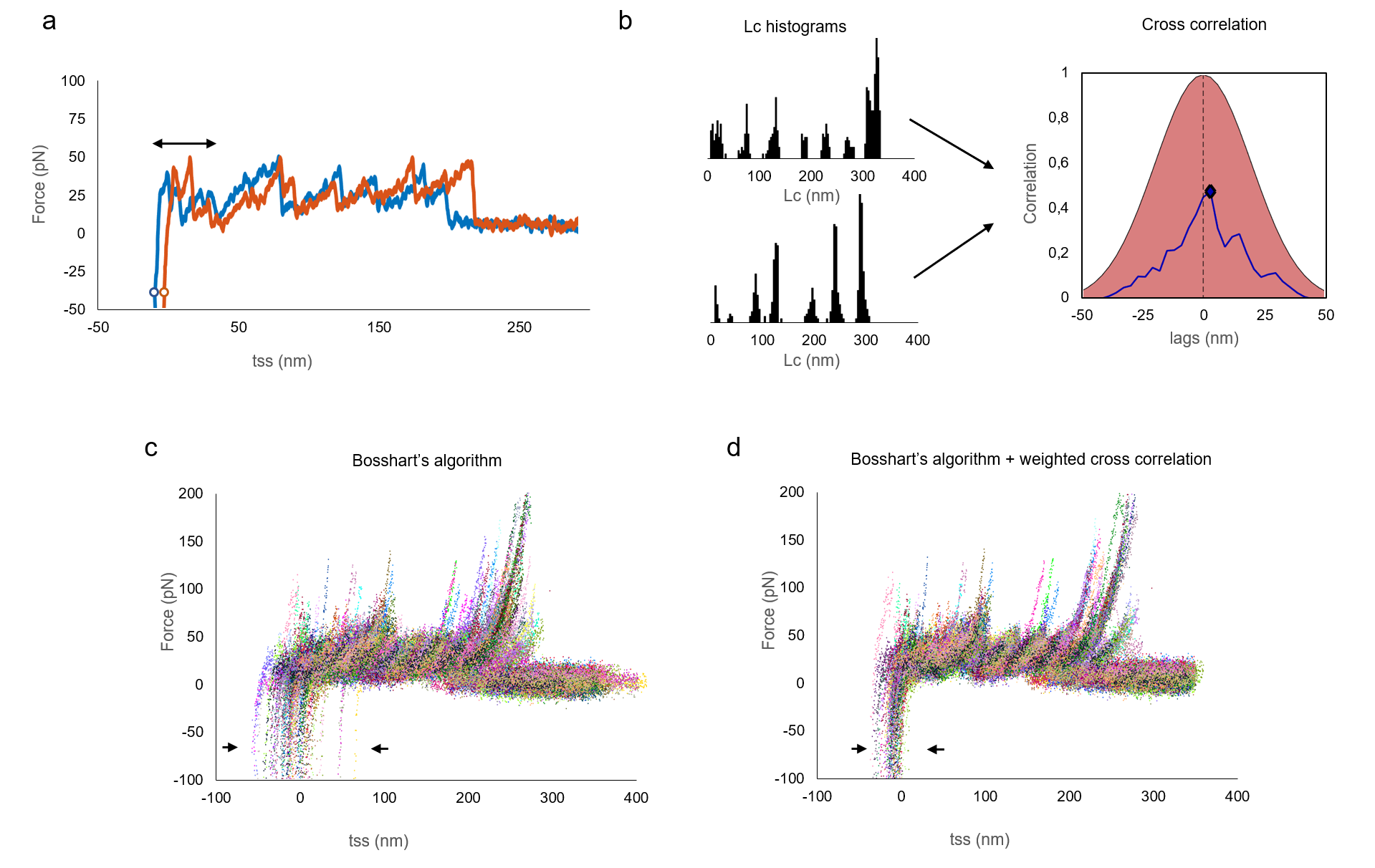}
	\captionsetup{justification=raggedright,singlelinecheck=false}.
	\makeatletter
	\renewcommand{\fnum@figure}{\figurename~S3}
	\makeatother
	\caption{\textbf{Automatic alignment }: (a) example of 2 misaligned curves. White circles indicate the zero of the of the tss (i.e. the z-position of the sample). (b) cross-correlation curve (blue line) between two contour length histograms. The maximum value of the correlation is highlighted by the blue diamond at lag=+2 nm; the vertical dotted line is positioned at 0 lag (i.e. where the zeros of the two curves coincide). The Gaussian distribution (red area) weights the cross-correlation curve (see Supplementary note 3, equation (3) and (4)). (c) superimposition of 106 curves automatically aligned with original Bosshart’s algorithm (8). This algorithm does not prevent curves to be shifted from the origin by excessive large values. (d) the same superimposition automatically aligned with Bosshart’s plus the weighted correlation curve (b) and in this way large shifts are reduced and almost eliminated.   }
	
\end{figure*}

\begin{figure*}
	
	\centering
	\includegraphics[width=0.6\linewidth]{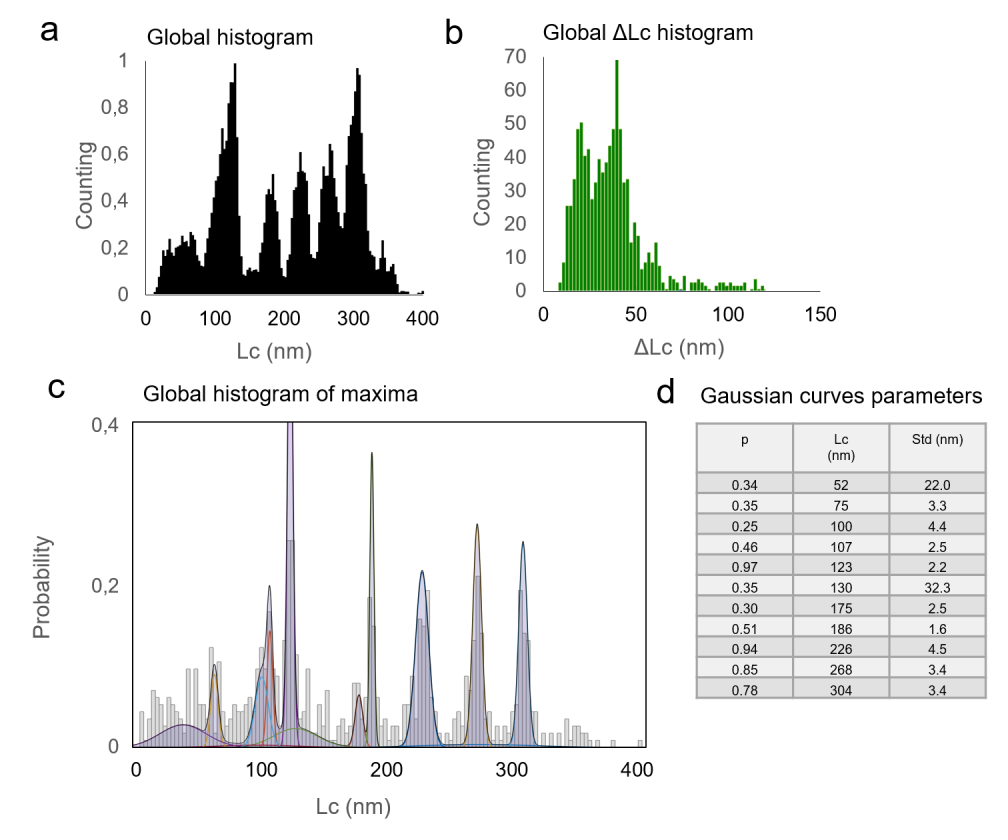}
	\captionsetup{justification=raggedright,singlelinecheck=false}.
	\makeatletter
	\renewcommand{\fnum@figure}{\figurename~S4}
	\makeatother
	\caption{\textbf{Analysis of the distribution of contour lengths (Lc)}:  Statistical analysis of selected curves. (a) The Global Histogram is the sum of all Lc histograms (Supp. Fig. 1d) normalized to the maximum bin value. The peak sharpness indicates a high homogeneity of F-d curves. (b) Global ΔLc histogram counts all increase of the values of Lc between two consecutive peaks. (c) Global histogram of maxima counts the values of the Lc of all detected peaks. The distribution can be fitted with multi Gaussians to determine the position and frequency of persistent unfolding events (see Supplementary Note 4 for the Gaussian mixture model). (d) probability, contour length (Lc) and standard deviation of the Gaussian curves displayed in (c).}
	
\end{figure*}

\begin{figure*}
	
	\centering
	\includegraphics[width=0.6\linewidth]{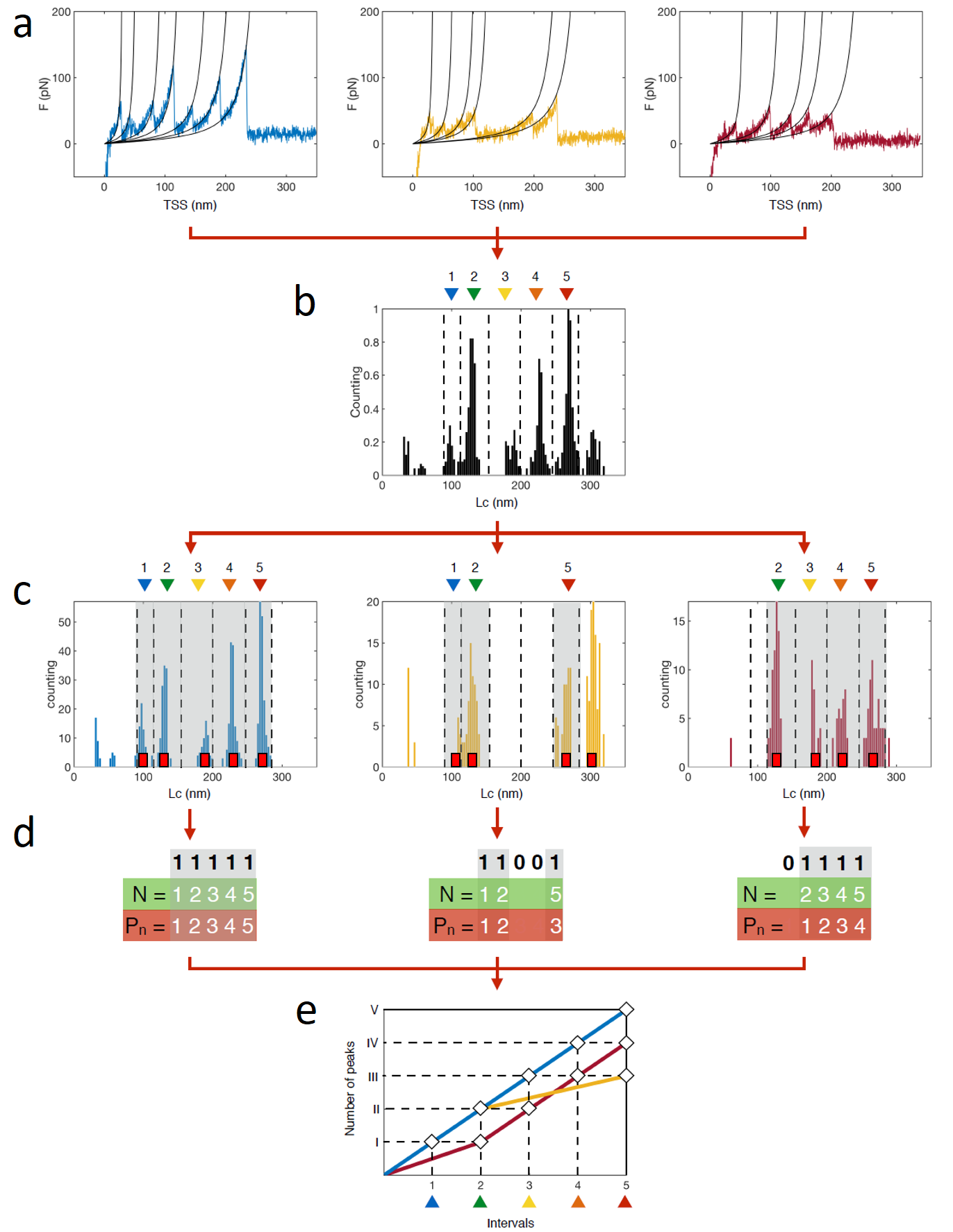}
	\captionsetup{justification=raggedright,singlelinecheck=false}.
	\makeatletter
	\renewcommand{\fnum@figure}{\figurename~S5}
	\makeatother
	\caption{\textbf{Pathplot construction}:  The Pathplot is generated in 5 steps and is used to find clusters of curves with force peaks with similar values of Lc. (a) Curves are fitted with the WLC model, transformed into Lc histograms and grouped in the Global Histogram as described in Figure S4 a. (b) The Global Histogram is divided in intervals according to the  ensemble of events so to obtain a partition of the Lc coordinates in regions with distinct maxima. (c) Following the aforementioned rule, in this explanatory panel, we selected 5 intervals corresponding to the intervals 90-118,118-155,155-200,200-247,247-285 nm. (d) On the basis of this division, each trace is coded in a binary string of 5 digits: 0 is assigned if no force peak is detected within the interval, 1 is assigned if at least one event is detected. From each string, we created two additional sequence: $\sharp$ and Pn. $\sharp$ is the sequence referred to the order of appearance of the force peak along the trace (in a trace with 2 peaks, the 1st peak has $\sharp$=1 and the 2nd peak has $\sharp$=2). Pn refers to the interval position occupied by a peak (a peak that fall within the 3rd interval has Pn=3). (e) Coded curves are plotted as broken lines into an orthogonal $\sharp$-Pn space, the line-width is proportional to the number of curves that follow the same path. This algorithm generalizes the previous proposed methods (9, 10) providing a representation able to distinguish different unfolding behaviors/clusters of a given set of curves, based on the number and position of occurrence of unfolding events. }
	
\end{figure*}

\begin{figure*}
	
	\centering
	\includegraphics[width=0.7\linewidth]{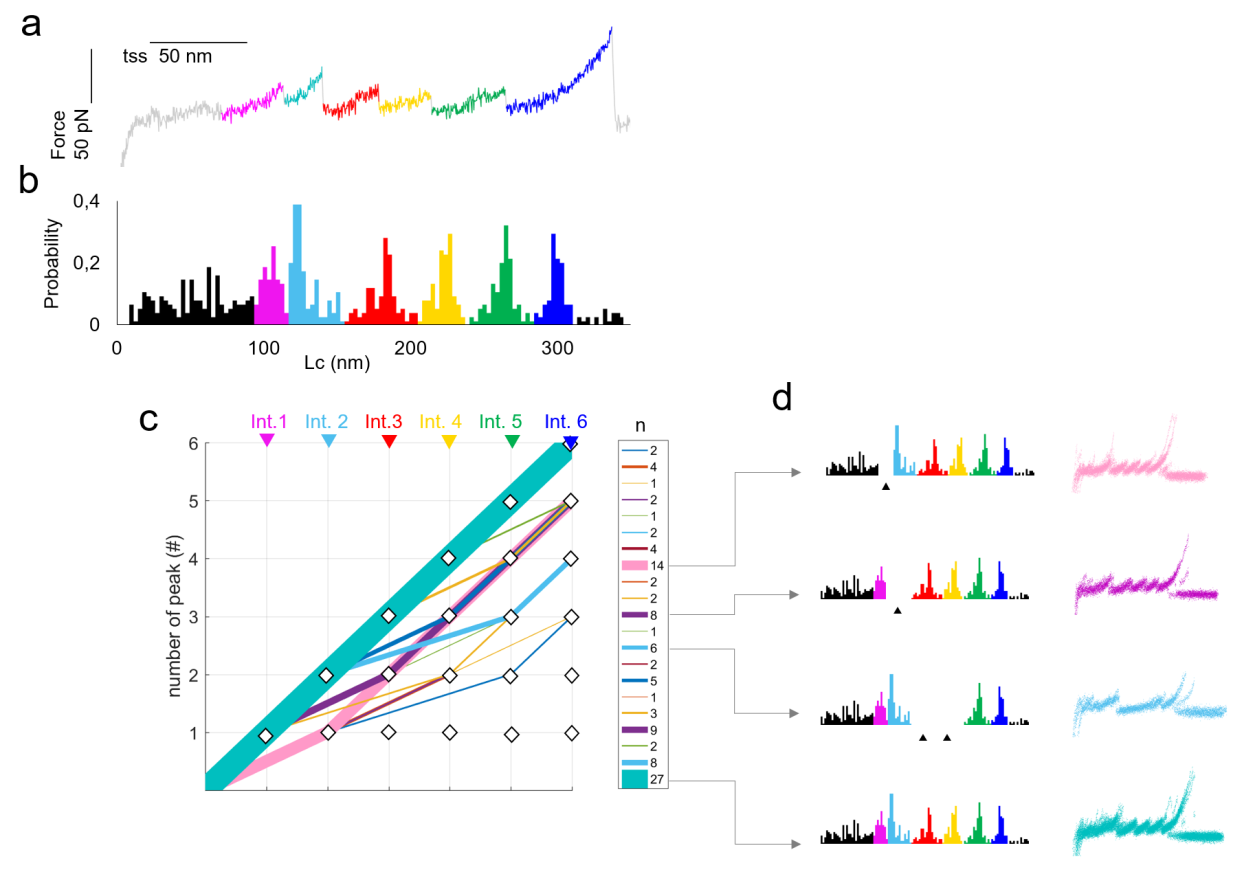}
	\captionsetup{justification=raggedright,singlelinecheck=false}.
	\makeatletter
	\renewcommand{\fnum@figure}{\figurename~S6}
	\makeatother
	\caption{\textbf{Pathplot of CNGA1 curves}:  (a) a F-d curve from the unfolding of CNGA1 channels (11). (b) Global Histogram of Maxima (Fig. S1 d) applied to the population of 106 curves of CNGA1 (Supplementary dataset). The major peaks of the histogram are highlighted with different colors. (c) the Pathplot applied to this set of curves with Lc intervals congruent with (b): 90-118,118-155,155-200,200-247,247-285,285-320 nm. The legend on the right shows the number of curves belonging to the different pathways. (d) examples of identified clusters with the corresponding set of F-d curves and the Lc histograms.	}
	
\end{figure*}

\balancecolsandclearpage

\section*{Supplementary notes}

\subsection*{Note 1}
\textit{Data Import}\\
Fodis has been designed to open raw data files from JPK and Bruker. To allow other users to operate with Fodis, but also to facilitate the portability of datasets, it can load (and export) data in ASCII format as explained in the User Guide section 1.2.\\

\textit{Open Samples} \\
The open samples option opens .jpk-force raw files and .txt files generated by JPK’s Data Processing software.
The open samples option opens .spm, .xyz (e.g. .084, .017 etc.) raw files and .txt generated by Bruker’s NanoScope (more details can be found in the User guide section 1.2.2). Support for other formats can be requested at the dedicated section of the web site https://github.com/nicolagalvanetto/Fodis/issues.\\

\textit{Import Curves}\\
Fodis imports .txt files of space separated numbers. The files should be organized in columns of Force and Distance values. F1 X1 F2 X2 F3 X3 … FN XN is the order of the columns, where F1 is the first trace Force values, X1 is the tip sample separation of the first trace and so forth until the N-th trace. For more details check the User Guide section 1.2.1 and the Supplementary data files.

\subsection*{Note 2}
\textit{Curves pre-processing, curves transformations and peaks detection}\\
\textit{-Pre-processing}\\
Force-distance curve in Fodis must have in the x-axis the Tip Sample Separation in meters (TSS) with the contact point on the left, and in the y-axis the force in Newton (baseline corrected).

The imported data may not be already formatted for the analysis (wrong units, TSS transformation needed, etc). The tool Absolute height to TSS has been developed for this scope. It can perform 6 different types of transformations:
\begin{itemize}
	\item Geometrical Adjustment: the mirroring option inverts the values of the Force. The rotation option swaps the coordinates.
	
	\item  X-axis Adjustment: the x-axis values are multiplied by the constant Height multiplier (which is automatically read in the import files when present). This is necessary when the imported files contain only the non-corrected Height channel (e.g. in JPK files)
	
	\item Y-axis Adjustment: if the vertical deflection of the cantilever is imported in Volts, the curves need to be rescaled with the Sensitivity constant and the Spring constant. If the vertical deflection of the cantilever is imported in meters, the curves need to be rescaled with the Spring constant. (The constants are automatically read in the imported files when present).
	
	\item Subtract Baseline: if the Force values need to be vertically shifted, this option finds the baseline averaging the final part of the curve (red patch in the tool). The starting point and the ending point of the patch can be changed in the edit boxes.
	
	\item Find contact point: it finds the first positive value.
	
	\item From piezo height to TSS: it transforms the x-coordinate in tip sample separation (TSS). (The constant is automatically read in the import files when present).

\end{itemize}

\textit{-WLC-based transformations}\\
F-d curves were first smoothed with Savitzky–Golay filter to reduce white noise (Fig. S1 b). Smoothed F-d curves were  converted into a plot of force and contour lengths (F, Lc) (3, 8) (Fig. S1 c). The Worm-like chain approximate equation used to perform curves transformation is (2):
\begin{equation}
	Fp/(k_b T)=1/4 (1-d/L_c )^{-2}-1/4+d/L_c  
\end{equation}

Here, the contour length is represented by $Lc$, the persistence length by $p$, the extension is represented by $d$, the external force is represented by $F$, and $k_b$ and $T$ are the Boltzmann constant and the absolute temperature. A persistence length (p) of 0.4 nm was used to compute the Lc value for each corresponding tip-sample separation (d or TSS) value, by solving the third order polynomial: 

\begin{equation}
4\lambda 3+ \omega \lambda 2-1=0
\end{equation}

where $\lambda= 1 - d/Lc  $ and $\omega = 4F(d,Lc)/\alpha - 3$ and $\alpha= k_b T/p$. This equation had three roots and the root of interest was the real root $\lambda $ such that $0<\lambda<1 $. In this manner, each point of the F-d curve (F, d) was transformed into a corresponding point (F, Lc), and each F-d curve was transformed into an (F, Lc) plot (Fig. S1 c). Owing to this transformation, each portion of the F-d curve that was fitted perfectly by a constant Lc in the WLC model was mapped to a perfect vertical segment. The transformation of an F-d curve was therefore a relation (set of point) in the (F, Lc) plane rather than a function in the plane, and it was also not a continuous curve. Given the F-Lc values, the histogram of the counts/bin of Lc values (Lc histogram) was computed (Fig. S1 d). The Lc axis of the (F, Lc) plot was first divided into bins. All points with a value of F larger than 30 pN were counted in the corresponding bin and summed. \\

-Peaks detection
The automatic detection of peaks (i.e. unfolding events) was done operating on the Contour length (Lc, Fc) plot (Fig. S1 c). A Force Profile is created dividing the Contour length (Lc, Fc) in bins, and taking the maximum force value in each bin. Then, Fodis uses the MATLAB function findpeaks( ) on the Force Profile, and it detects every maxima of the profile. Fodis allows to tune the Force Profile operating on the smoothing, on threshold N points (i.e. number of points to neglect for each bin) and the minimum peak proximity (see user guide, section 1.5.4). The default parameters are optimized to detect the force peaks of the CNG unfolding trace. A shorter F-d curve will require a smaller minimum peak proximity. Binary strings containing the information of peaks position are plotted as red bars in Lc histogram view (Fig. S1 d) and are grouped together in the Global Matrix (Fig. 1 e of the main text).

\subsection*{Note 3}
Cross-correlation and automatic alignment \\
F-d curves to be compared to each other may require horizontal alignment. The main cause of the lateral shift are the different attachment points between the protein terminus and the tip, but also other effects may be present. Bosshart and colleagues (8) developed a reference free alignment method consisting of 4 steps. Starting from the contour length histogram of every trace (Fig. S1 d), they: 
\begin{itemize}
	\item subdivided the curves into groups of homogeneous curves (i.e. curves with the same number of peaks);
	
	\item recursively aligned curves into the same group with the maximum correlation principle, building an average contour-length reference for each group;
	
	\item formed a global reference (Fig. S3 c);
	
	\item aligned all the curves of the dataset to the global reference.

\end{itemize}

This procedure is suitable for identical globular multidomain proteins where force peaks generally occurred within conserved Lc values. Instead, in the case of soluble and complex proteins or membrane proteins, the occurrence of unfolding events may be variable due to the stochastic nature of the process, or because of the multiplicity of the unfolding pathways that may be accessed by the protein. For this reason, we introduced two additional features to their procedure:
\begin{enumerate}
	\item In addition to the contour-length histogram, we assign to every trace a zero-point, that is the point of tip-sample contact. Given the correlation curve (C; equation 3) of two curves, we then multiply the correlation curve with a Gaussian curve centered at the point in which the zero-points of the two curves match with each other (equation 4; Fig. S3 b). The idea is to apply a “potential well” to reduce the maximum displacement of the two zeros.
	
	\item  Group division proposed by Bosshart and colleagues works only if all the curves with the same number of peaks have the peaks in the same position, but this is not generally true for F-d curves of the same protein. Therefore, we used a group division following the method described in Fig S5. In this way, we imposed two constrains for a given trace to be part of a given group: to have a speciﬁc number of peaks and to have these peaks in a speciﬁc position.
\end{enumerate}

%
%

\subsection*{Note 4}
\textit{Multi Gaussian Fitting of Global Histogram of Maxima (GHM)}\\
GHM shows the distribution of unfolding events (peaks) along the Lc coordinate. In 2013 it has been shown by Kawamura and colleagues (12) that an ideal GHM can be fitted by a Gaussian mixture model to determine the probability of occurrence of a certain unfolding event (likely corresponding to a stable structural segment). The fitting implemented in Fodis uses fitgmdis MATLAB function over 100 iterations. To determine the correct number of Gaussian bells that best fit the distribution, we computed N different distribution of Gaussian mixtures {g1, … , gN} where g1 has 1 Gaussian bell, and gN has N Gaussian bells. The best distribution gX is the one that minimize the Akaike Information Criterion (AIC) (13). A real GHM is characterized by a distribution that is the sum of Gaussian bells, plus a constant background noise (12). To overcome this problem, Fodis divides the Lc values relatively to the selected bin size, and within each bin it randomly removes one point of Lc distribution. In this way, by setting different bin sizes it is possible to tune the background noise and enhance peak detection and fitting.

\bibliography{bib}

\subsection*{Supplementary References}

1. 	Savitzky, A., and M.J.E. Golay. 1964. Smoothing and Differentiation of Data by Simplified Least Squares Procedures. Anal. Chem. 36: 1627–1639.

2. 	Marko, J.F., and E.D. Siggia. 1995. Statistical mechanics of supercoiled DNA. Phys. Rev. E. 52: 2912–2938.
3. 	Carrion-Vazquez, M., A.F. Oberhauser, T.E. Fisher, P.E. Marszalek, H. Li, and J.M. Fernandez. 2000. Mechanical design of proteins studied by single-molecule force spectroscopy and protein engineering. Progress in Biophysics and Molecular Biology. 74: 63–91.

4. 	Oesterhelt, F., D. Oesterhelt, M. Pfeiffer, A. Engel, H.E. Gaub, and D.J. Müller. 2000. Unfolding Pathways of Individual Bacteriorhodopsins. Fodis. 288: 143–146.

5. 	Celik, E., and V.T. Moy. 2012. Nonspecific interactions in AFM force spectroscopy measurements. J. Mol. Recognit. 25: 53–56.

6. 	Amestoy, P., T. Davis, and I. Duff. 1996. An Approximate Minimum Degree Ordering Algorithm. SIAM. J. Matrix Anal. \& Appl. 17: 886–905.

7. 	Kawamura, S., M. Gerstung, A.T. Colozo, J. Helenius, A. Maeda, N. Beerenwinkel, P.S.-H. Park, and D.J. Müller. 2013. Kinetic, Energetic, and Mechanical Differences between Dark-State Rhodopsin and Opsin. Structure. 21: 426–437.

8. 	Akaike, H. 1974. A new look at the statistical model identification. IEEE Transactions on Automatic Control. 19: 716–723.

\end{document}